\documentstyle[aps,prd,amssymb,eqsecnum]{revtex}

\newcommand{\be}{\begin{equation}}
\newcommand{\ee}{\end{equation}}
\newcommand{\bea}{\begin{eqnarray}}
\newcommand{\eea}{\end{eqnarray}}
\newcommand{\ben}{\begin{eqnarray*}}
\newcommand{\een}{\end{eqnarray*}}

\def\np{({\bf n}\cdot{\bf p})}
\def\pp{{\bf p}^2}

\def\npi{({\bf n}_{12}\cdot{\bf p}_1)}
\def\pipi{{\bf p}_1^2}
\def\pipip{({\bf p}_1^2)}

\def\npj{({\bf n}_{12}\cdot{\bf p}_2)}
\def\pjpj{{\bf p}_2^2}
\def\pjpjp{({\bf p}_2^2)}

\def\pipj{({\bf p}_1\cdot{\bf p}_2)}

\def\omk{\omega_{\text{kinetic}}}
\def\oms{\omega_{\text{static}}}

\def\hn{H_{\text{N}}}
\def\hi{H_{\text{1PN}}}
\def\hii{H_{\text{2PN}}}
\def\hiii{H_{\text{3PN}}}

\def\nvi{({\bf n}_{12}^{\text{h}}\cdot{\bf v}_1)}
\def\nvj{({\bf n}_{12}^{\text{h}}\cdot{\bf v}_2)}

\def\vivi{{\bf v}_1^2}
\def\vivip{({\bf v}_1^2)}

\def\vjvj{{\bf v}_2^2}
\def\vjvjp{({\bf v}_2^2)}

\def\vivj{({\bf v}_1\cdot{\bf v}_2)}

\def\rh{r_{12}^{\text{h}}}
\def\brh{{\bf r}_{12}^{\text{h}}}
\def\nh{{\bf n}_{12}^{\text{h}}}

\def\nai{({\bf n}_{12}^{\text{h}}\cdot{\bf a}_1)}
\def\naj{({\bf n}_{12}^{\text{h}}\cdot{\bf a}_2)}
\def\viai{({\bf v}_1\cdot{\bf a}_1)}
\def\viaj{({\bf v}_1\cdot{\bf a}_2)}
\def\vjai{({\bf v}_2\cdot{\bf a}_1)}
\def\vjaj{({\bf v}_2\cdot{\bf a}_2)}

\def\tr{{\text{reg}}}

\begin{document}

\title{Equivalence between the ADM-Hamiltonian and the 
harmonic-coordinates approaches to the third post-Newtonian 
dynamics of compact binaries}

\author{Thibault Damour}
\address{Institut des Hautes \'Etudes Scientifiques,
91440 Bures-sur-Yvette, France}

\author{Piotr Jaranowski}
\address{Institute of Theoretical Physics,
University of Bia{\l}ystok,
Lipowa 41, 15-424 Bia{\l}ystok, Poland}

\author{Gerhard Sch\"afer}
\address{Theoretisch-Physikalisches Institut,
Friedrich-Schiller-Universit\"at,
Max-Wien-Pl.\ 1, 07743 Jena, Germany}

\maketitle

\begin{abstract}
The third post-Newtonian approximation to the general relativistic dynamics of
two point-mass systems has been recently derived by two independent groups,
using different approaches, and different coordinate systems.  By explicitly
exhibiting the map between the variables used in the two approaches we prove
their physical equivalence.  Our map allows one to transfer all the known
results of the Arnowitt-Deser-Misner (ADM) approach to the harmonic-coordinates
one:  in particular, it gives the value of the harmonic-coordinates Lagrangian,
and the expression of the ten conserved quantities associated to global
Poincar\'e invariance.

\vspace{2ex}\noindent
PACS number(s): 04.25.Nx, 04.20.Fy, 04.30.Db, 97.60.Jd
\end{abstract}

\section{Motivation}

Binary systems made of compact objects (neutron stars or black holes) are the
most promising sources for the upcoming ground-based network of interferometric
gravitational wave detectors LIGO/VIRGO/GEO.  Because of their higher
signal-to-noise ratio, the first detections are likely to involve massive binary
black-hole systems, with total mass $m_1+m_2\agt30\,M_{\odot}$.  Such systems
emit most of their useful signal at the end of their inspiral phase, near the
last stable (circular) orbit.  This makes it very important to have the best
possible analytical control of the general relativistic dynamics of two-body
systems.

For many years the equations of motion of binary systems have been known only up
to the 5/2 post-Newtonian (2.5PN) approximation
\cite{DD,D83,DS85,S85,S86,KG85-86,BFP98}.  Recently, Jaranowski and Sch\"afer
\cite{JS98,JS99} and Damour, Jaranowski, and Sch\"afer \cite{DJS1,DJS3}
succeeded in deriving the third post-Newtonian (3PN) dynamics of binary
point-mass systems within the canonical formalism of Arnowitt, Deser, and Misner
(ADM).  More recently, Blanchet and Faye \cite{BF1,BF4} succeeded in deriving
the 3PN equations of motion of binary point-mass systems in harmonic coordinates
relying on an independent framework.  The purpose of this paper is to compare
and relate these two sets of results.

For the present investigation it is sufficient to consider the conservative part
of the dynamics, i.e.\ we shall drop the dissipative 2.5PN part which is the
leading order radiation-damping level.  This has the advantage that for the
remaining part there exists an autonomous Hamiltonian (depending only on
particle variables) and a conserved energy.  The non-autonomous parts of the ADM
Hamiltonian, up to the 3.5PN level (which is the next order radiation-damping
level after 2.5PN), are given in Ref.\ \cite{JS97}.

\section{Regularization ambiguities}

Before tackling the comparison between the two sets of 3PN results several
remarks are in order.  First, let us emphasize that both approaches to the 3PN
dynamics have found that the use of Dirac-delta-function sources to model the
two-body system causes the appearance of both badly divergent integrals and
badly defined ``contact terms'', which (contrary to what happened at the 2.5PN
\cite{D83,S85} and 3.5PN \cite{JS97} levels) cannot be unambiguously
regularized.  More precisely, when Refs.\ \cite{JS98,JS99} derived the
relative-motion 3PN ADM Hamiltonian $H({\bf x},{\bf p})$, in the center-of-mass
frame of the binary, they introduced two arbitrary {\em dimensionless}
parameters, $\omega_k(\equiv\omk$) and $\omega_s(\equiv\oms$), to formalize the
presence of irreducible ambiguities in the regularization of the Hamiltonian.
The regularization ambiguity parameter $\omega_k$ concerned a momentum-dependent
contribution $\propto G^3 c^{-6} (\pp-3\np^2)r^{-3}$, while $\omega_s$
concerned a momentum-independent contribution $\propto G^4c^{-6}r^{-4}$.  Ref.\
\cite{DJS3}, on the one hand generalized the work of \cite{JS98} by deriving the
3PN ADM Hamiltonian $H({\bf x}_1,{\bf x}_2,{\bf p}_1,{\bf p}_2)$ in an arbitrary
reference frame, and, on the other hand, proved that $\omega_k$ was uniquely
determined to have the value $\omega_k=41/24$ by requiring the global
Poincar\'e invariance of the 3PN dynamics (see Ref.\ \cite{JSWeimar} for details 
of why $\omega_k$ is not fixable in the center-of-mass frame).  Therefore, 
finally, the 3PN ADM Hamiltonian\footnote{
Note that we are considering here the ordinary 3PN Hamiltonian, obtained 
(following a result of \cite{DJS1}) by a well-defined shift of phase-space 
coordinates, designed to {\em reduce} the {\em higher-order} Hamiltonian 
$\widetilde{H}_{\text{3PN}}({\bf x}_a,{\bf p}_a,\dot{\bf x}_a,\dot{\bf p}_a)$ 
defined by eliminating the field variables $h^{\text{TT}}_{ij}$, 
$\dot{h}^{\text{TT}}_{ij}$ in the ``Routh functional''$R_{\text{3PN}}
({\bf x}_a,{\bf p}_a,h^{\text{TT}}_{ij},\dot{h}^{\text{TT}}_{ij})$ of 
\cite{JS98}.}
contains only {\em one} regularization ambiguity (parametrized by the {\em 
dimensionless} parameter $\omega_s$):
\begin{eqnarray}
\label{eq1}
H({\bf x}_a,p_a;\omega_s) &=& \sum_a m_ac^2 + \hn({\bf x}_a,p_a) 
\nonumber\\[2ex]&&
+ \frac{1}{c^2} \hi({\bf x}_a,{\bf p}_a)
+ \frac{1}{c^4} \hii({\bf x}_a,{\bf p}_a)
\nonumber\\[2ex]&&
+ \frac{1}{c^6} \left[ H^{\omega_s=0}_{\text{3PN}}({\bf x}_a,{\bf p}_a) 
+ \omega_s \frac{G^4m_1^2m_2^2(m_1+m_2)}{r_{12}^4} \right] \, . 
\end{eqnarray}
See Ref.\ \cite{DJS3} for the explicit expressions of $\hiii^{\omega_s=0}$, and 
of the well known lower-order (Newtonian, 1PN, and 2PN) contributions. 
  
On the other hand, Ref.\ \cite{BF4}, in deriving the 3PN equations of motion in
harmonic coordinates, introduced four arbitrary {\em dimensionfull} parameters
$s_1$, $s_2$, $r'_1$, and $r'_2$ (with dimensions of length; $s_1$, $s_2$
correspond to the intermediate ADM regularization length scales $l_1$, $l_2$
discussed below, which disappear in the final $\hiii$).  In addition, though
these authors developed some formal generalization of the theory of
distributions to deal with the badly divergent integrals appearing at 3PN
\cite{BF2}, they could not prove the uniqueness of their prescriptions, and, in
fact, they used two different prescriptions, the most recent of them \cite{BF4}
introducing a new {\em dimensionless} arbitrary parameter $K$.
(However, they did prove that both versions of their regularization 
prescriptions finally lead to gauge-equivalent equations of motion.)
In \cite{BF3} these authors also introduced a modification of their 
regularization procedures, aimed at yielding (``in principle'') 
Lorentz-invariant equations of motion.  The net result of using the set 
of regularization recipes developed in \cite{BF2,BF3,BF4} is the derivation of 
3PN two-body equations of motion, in harmonic coordinates, which depend on the 
{\em five} parameters $s_1$, $s_2$, $r'_1$, $r'_2$, and $K$ and which are, 
generically, neither Lorentz-invariant,\footnote{At least if one follows 
\cite{BF4} in using the new ``correct'' derivative involving the parameter $K$.}
nor deducible from an action (because they do not lead to a conserved
energy as any autonomous action-based equations of motion would).  Then, the
authors of \cite{BF4} impose the triple requirement of: (i) Lorentz invariance,
(ii) existence of a conserved energy, and (iii) polynomiality in $m_1$ and
$m_2$.  They show that:  (i) uniquely determines $K$ to have the value
$K=41/160$, and (ii) imposes one constraint relating the four length scales
$s_1$, $s_2$, $r'_1$, and $r'_2$, namely
\be
\label{eq2}
m_2 \left[ \ln\left(\frac{r'_2}{s_2}\right) +  \frac{783}{3080} \right]
=  m_1 \left[ \ln\left(\frac{r'_1}{s_1}\right) + \frac{783}{3080} \right].
\ee
Note that, when they use their older version of their regularization
prescriptions, the rational number appearing in Eq.\ (\ref{eq2}) becomes
$-159/308$.  By further imposing the requirement (iii), they conclude that the
two length scales $s_1,s_2$ can be expressed in terms of the two other scales
$r'_1$, $r'_2$, and of a new {\em dimensionless} parameter $\lambda$, through
\begin{mathletters}
\begin{eqnarray}
\label{eq3}
\ln\left(\frac{r'_1}{s_1}\right) &=& -\frac{783}{3080}
+ \lambda \frac{m_1+m_2}{m_1}, 
\\[2ex]
\ln\left(\frac{r'_2}{s_2}\right) &=& -\frac{783}{3080}
+ \lambda \frac{m_1+m_2}{m_2}.
\end{eqnarray}
\end{mathletters}

Finally, the 3PN equations of motion for the harmonic coordinates ${\bf y}_a(t)$ 
of the two point masses  contain {\em three} regularization ambiguities 
(parametrized by the two scales $r'_1$, $r'_2$ and the dimensionless parameter 
$\lambda$) and have the form $\ddot{{\bf y}}_a={\bf A}_a({\bf y}_b,{\bf v}_b)$, 
where ${\bf v}_a\equiv\dot{{\bf y}}_a$, with
\begin{eqnarray}
\label{eq4}
{\bf A}_a({\bf y}_b,{\bf v}_b) &=&
{\bf A}_{a\,{\text{N}}}({\bf y}_b,{\bf v}_b)
+ \frac{1}{c^2} {\bf A}_{a\,{\text{1PN}}}({\bf y}_b,{\bf v}_b)
+ \frac{1}{c^4} {\bf A}_{a\,{\text{2PN}}}({\bf y}_b,{\bf v}_b)
+ \frac{1}{c^5} {\bf A}_{a\,{\text{2.5PN}}}({\bf y}_b,{\bf v}_b)
\nonumber\\[2ex]&&
+ \frac{1}{c^6} \left[
{\bf A}_{a\,{\text{3PN}}}^{(0)}({\bf y}_b,{\bf v}_b)
+ \ln\left(\frac{\rh}{r'_1}\right)
{\bf A}_{a\,{\text{3PN}}}^{(1)}(\brh,{\bf v}_{12})
+ \ln\left(\frac{\rh}{r'_2}\right)
{\bf A}_{a\,{\text{3PN}}}^{(2)}(\brh,{\bf v}_{12}) +
\lambda\, {\bf A}_{a\,{\text{3PN}}}^{(3)}(\brh) \right],
\end{eqnarray}
where $\brh\equiv{{\bf y}_1-{\bf y}_2}$ and
${\bf v_{12}}\equiv\dot{\bf r}^{\text{h}}_{12}\equiv{\bf v}_1-{\bf v}_2$
denote the (harmonic) relative position and velocity, respectively.
It was, however, shown in \cite{BF4} that the ambiguities linked to $r'_1$, 
$r'_2$ can be gauged away, so that the physical ambiguity of the harmonic 
equations of motions is described by the {\em sole} (dimensionless) parameter 
$\lambda$.
For simplicity, we shall work here with the equations of motion explicitly
displayed in \cite{BF4} which, in fact, corresponds to their older
regularization prescription [with 783/3080 being replaced by $-159/308$ in Eq.\
(\ref{eq2})].  See Eq.\ (7.16) of \cite{BF4} for the explicit expression of the
3PN contributions to the harmonic equations of motion (as well as of the well
known lower-order contributions \cite{DD,D83}).  We shall only note here the
fact that
$A_{a\,{\text{3PN}}}^{(1)}$ and $A_{a\,{\text{3PN}}}^{(2)}$ depend only on the 
relative positions and velocities, and that the $\lambda$-term reads (for $a=1$; 
$\nh\equiv\brh/\rh$)
\begin{eqnarray}
\label{eq5}
\lambda\, {\bf A}_{1\,{\text{3PN}}}^{(3)}(\brh)
= -\frac{44}{3} \lambda \frac{G^4m_1m_2^2(m_1+m_2)}{(\rh)^5} \nh \, .
\end{eqnarray}
Even before any detailed calculation, it is clear that this 
$\lambda$-contribution derives from a potential energy 
$$
\lambda\, V^{(3)} \equiv - \frac{11}{3} \lambda
\frac{G^4 m_1^2 m_2^2 (m_1+m_2)}{c^6 (\rh)^4},
$$
so that, if the two different 3PN dynamics can be shown to be somehow 
equivalent, the ``harmonic'' regularization ambiguity $\lambda$ must be related 
to the ``ADM'' one $\omega_s$ by $-\frac{11}{3}\lambda=\omega_s$ + const.

\section{Origin of regularization ambiguities}

As the presence of the regularization ambiguities at the 3PN level is very
striking,\footnote{
Though it was anticipated in \cite{D83}, see pp.\ 107 and 116 there.}
and physically momentous, let us discuss in more detail the origin of the
ambiguities present in the two approaches, and the differences between them.

In the ADM approach, one computes a (spatially) {\em global scalar} quantity, 
the Hamiltonian $H({\bf x}_a,{\bf p}_a)$ of the system. Essentially\footnote{
After applying the double ``reduction'' process of eliminating the field 
variables and reducing the order of the Hamiltonian \cite{JS98,DJS1}.} 
the global scalar $H({\bf x}_a,{\bf p}_a)$ can be explicitly expressed as an 
integral over space of an integrand of the generic form
\be
\label{eq3.1}
{\cal H}({\bf x};{\bf x}_a,{\bf p}_a)
= {\cal H}^{(D)}_c({\bf x};{\bf x}_a,{\bf p}_a)
+ {\cal H}^{(D)}_f({\bf x};{\bf x}_a,{\bf p}_a)
+ \partial_iD^i({\bf x};{\bf x}_a,{\bf p}_a).
\ee
Here, ${\cal H}^{(D)}_c$ is made only of ``contact terms'', i.e.\ of terms
proportional to the delta-functions modelling the sources, say
${\cal H}^{(D)}_c=\sum_a S_a({\bf x},{\bf x}_b)\delta({\bf x}-{\bf x}_a)$, where 
$S_a$ is constructed from field quantities, ${\cal H}^{(D)}_f$ is a
``field-like'' term, i.e.\ an ``energy density'' constructed from field
quantities and distributed all over space, and the last term is a pure
divergence, which formally gives a vanishing contribution\footnote{
It has been checked in the ADM approach that the ``surface term at infinity''
associated to $\partial_iD^i$ is not causing any ambiguity.  Indeed, most pieces
in $\oint dS_iD^i$ decay like some inverse power of $r$ at infinity, while the
ones which might be problematic (like the one associated to the ${\cal O}(r)$
part of $h^{\text{TT}}_{(6)ij}$) have been explicitly shown to give a vanishing
contribution to $\oint dS_iD^i$.  The ambiguities come only from the singular
behaviour of the integrand near each particle, i.e.\ as the field point
${\bf x}$ tends to either ${\bf x}_1$ or ${\bf x}_2$.}
to the integrated Hamiltonian.  As
indicated by the superscript notation, the explicit values of the ``contact''
and ``field'' terms depend on the choice of the divergence term $\partial_iD^i$.
In other words, we can, by ``operating by parts'', shuffle terms between ${\cal
H}^{(D)}_c$ and ${\cal H}^{(D)}_f$, at the price of changing $D^i$.  Note that,
when so shuffling terms, one freely uses Einstein's field equations (with
delta-function sources) and one assumes the validity of the usual rules of
functional calculus,\footnote{
In the explicit computations of the Hamiltonian done in Refs.\ \cite{JS98},
\cite{DJS1}, and \cite{DJS3} one has always chosen $D^i$'s such that all the
terms in ${\cal H}^{(D)}_f$ contain only one derivative (or its equivalent)
acting on the elementary fields ($\phi_{(2)}$, $\pi^i_{(3)}$, $\phi_{(4)}$,
\ldots), so that there is no need to worry about using an improved
distributional derivative.  The distributional rule of differentiation of
homogeneous functions described in Appendix B of \cite{JS98} is, in fact, used
only when gauging the ambiguities by computing the regularized value of $\int
d^3x(\partial_iD^i)$, as explained in the Appendix A of \cite{DJS1}.}
such as Leibniz' rule [$\partial_i(AB)=(\partial_iA)B+A(\partial_iB)]$, and the 
commutativity of repeated derivatives 
($\partial_i\partial_jA=\partial_j\partial_iA$).

The ambiguities in the determination of the value of
$H\equiv\int{d^3x\,{\cal H}}$ come from two separate (but related) facts. First, 
the ``contact'' contribution
$$
H^{(D)}_c \equiv \int d^3x\, {\cal H}^{(D)}_c
= \int d^3{\bf x} \sum_a S_a({\bf x},{\bf x}_b)\delta({\bf x}-{\bf x}_a)
$$
is formally infinite because the (field-constructed) quantity
$S_a({\bf x},{\bf x}_b)$ is generically singular as ${\bf x}\to{\bf x}_a$.  To 
give a meaning to $H^{(D)}_c$ one must choose a specific regularization 
prescription to define the limit
$\lim_{{\bf x}\to{\bf x}_a}S_a({\bf x},{\bf x}_b)$.  Second, the ``field'' 
contribution $H^{(D)}_f \equiv \int d^3x {\cal H}^{(D)}_f$ is also formally 
infinite because the integrand ${\cal H}^{(D)}_f$ is generically too singular as 
${\bf x}\to{\bf x}_a$ to be locally integrable.  To give a meaning to 
$H^{(D)}_f$ one 
must choose a specific regularization prescription for such singular integrals.  
Finally for each choice of $D^i$, one {\em defines} the regularized value of the 
Hamiltonian as $H^{(D)\tr} \equiv H^{(D)\tr}_c + H^{(D)\tr}_f$.

In Refs.\ \cite{JS98,JS99} the following specific regularization prescriptions 
were adopted:
(i) for contact terms
$[\lim_{{\bf x}\to {\bf x}_a} S_a({\bf x},{\bf x}_b)]^\tr$ is defined as 
Hadamard's ``partie finie'' of $S_a({\bf x},{\bf x}_b)$,
Pf$_a S_a({\bf x},{\bf x}_b)$, defined in Appendix B of \cite{JS98} as the 
angle-averaged finite term in the Laurent expansion of
$S_a({\bf x}_a+{\bf r}_a,{\bf x}_b)$ in powers of
$r_a\equiv{\vert{\bf r}_a\vert}\equiv{\vert{\bf x}-{\bf x}_a\vert}$
(as ${\bf r}_a\to0$),
and (ii) for field terms the regularized value $I^\tr$ of a singular integral
$I=\int d^3x F({\bf x},{\bf x}_1,{\bf x}_2)$ is defined as
follows. First, one regularizes separately the divergences near each particle,
i.e.\ the integrals $I_a\equiv\int_{V_a} d^3x F({\bf x},{\bf x}_1,{\bf x}_2)$ 
where $V_a$ is a volume which contains ${\bf x}_a$ but not ${\bf x}_b$, with 
$b\neq a$.
[Evidently, one can always decompose $I=I_1+I_2+I_{\text{compl}}$ with two local 
volumes $V_1$, $V_2$ and a regular complement.]
Second, each local integral, say $I_1$ near particle 1, is
regularized ``{\`a} la Riesz'', i.e.\ by analytic continuation (AC) in 
$\epsilon_1$ of
$$
I_1(\epsilon_1)\equiv\int_{V_1} d^3x
\left(\frac{r_1}{l_1}\right)^{\epsilon_1} F({\bf x},{\bf x}_1,{\bf x}_2),
$$
where $r_1=\vert{\bf x}-{\bf x}_1\vert$ and where $l_1$ is a certain length
scale.\footnote{
The ``Riesz'' prescription explained in the Appendix of B of \cite{JS98} looks 
different from what we explain here (because it does not separate the 
integration volume in $V_1$, $V_2$ and the rest), but, as emphasized in 
\cite{DJS1}, it is equivalent to the logically simpler prescription that we 
summarize here.}
Most integrands $F$ lead to functions of $\epsilon_1$, $I_1(\epsilon_1)$, which 
are analytically continuable into the complex $\epsilon_1$-plane down to 
$\epsilon_1=0$.  In such a case this continuation 
AC$_{\epsilon_1\to0}I_1(\epsilon_1)$ uniquely defines the regularized value of 
$I_1$.  However, a limited subclass of ``dangerous''integrals gives rise to a 
(simple) pole as $\epsilon_1 \to 0$:  
$I_1(\epsilon_1)=Z_1\biglb(\epsilon_1^{-1}+\ln(R_1/l_1)\bigrb)+A_1$, where $R_1$ 
is an ``infra-red'' length scale associated to the choice of the local volume 
$V_1$.  For such integrals, one is naturally led (following the usual ``minimal 
subtraction'' prescription of quantum field theory) to defining the regularized 
value of $I_1(\epsilon_1)$ as the limit of $I_1(\epsilon_1)-Z_1/\epsilon_1$ as
$\epsilon_1\to0$, i.e.\ as $I_1^\tr\equiv{Z_1\ln(R_1/l_1)+A_1}$.  Note
that this regularization prescription has introduced one arbitrary length scale:
the regularization length $l_1$.  [The $V_1$-related infra-red length $R_1$ is
easily seen to cancel out in $I=I_1+I_2+I_{\text{compl}}$.]  However, as 
emphasized in Sec.\ IV of \cite{JS98}, a remarkable thing occurs in the explicit
calculations of the 3PN ADM Hamiltonian:  the combination of dangerous integrals
appearing in $H_{3PN}$ is such that all pole terms {\em exactly cancell}:
$\sum{Z_1}=0$.
In fact, one of the characteristics of the calculation of $\hiii$ in the ADM 
formalism is that one finds it much safer (and simpler) to regularize, 
at once, the full integral, rather than to try (as in the harmonic-coordinate 
calculation, \cite{BF4}) to give a separate regularized value for each 
individual contribution to the equations of motion.  The global
cancellation of the poles shows that the combination of dangerous integrals
appearing in $\hiii$ is of a less dangerous type.  A nice aspect of this
cancellation is that, in the ADM approach, the two regularizing length scales
$l_1$, $l_2$ completely cancell and do not appear in the regularized final
$\hiii$.  This does not mean, however, that the final result is unambiguous.
Indeed, it was emphasized in \cite{JS98,DJS1} that the regularized value of
$\hiii$ depends on the reshuffling of terms used to separate $\cal{H}$ in ${\cal
H}^{(D)}_c+{\cal H}^{(D)}_f+\partial_iD^i$.  In other words, when operating by
parts (which changes $D^i$, and ${\cal H}^{(D)}_c$ and ${\cal H}^{(D)}_f$) the
regularized value of $H^{(D)} \equiv H^{(D)}_c + H^{(D)}_f$ is found to
change.\footnote{
In actual calculations (see especially Appendix A of \cite{DJS1}) one monitors 
the changes in $H^{(D)} \equiv H^{(D)}_c + H^{(D)}_f$ by computing the 
term-by-term {\em regularized} value of the full algebraic expansion of the 
divergence term $\int d^3x(\partial_iD^i)$.}
In addition to this $D^i$-dependent ambiguity, there is also the problem of the 
sensitivity of the contact contribution $H^{(D)}_c$ to the choice of 
prescription for defining the `partie finie' of $S_a({\bf x},{\bf x}_b)$.  It 
was emphasized in \cite{JS99,DJS1} that the definition of the `Hadamard partie 
finie' Pf$_a$ becomes ambiguous at 3PN because it cannot be `threaded' through a 
product of field functions, i.e.\ that, in general,
Pf$_a(f_1f_2\ldots)\neq($Pf$_af_1)($Pf$_af_2)\ldots$
[The prime, irreducible example of this ambiguity at 3PN comes from the fact 
that Pf$_a(\phi_{(2)}^4)\neq[$Pf$_a(\phi_{(2)}^2)]^2=[$Pf$_a(\phi_{(2)})]^4$,
where $\phi_{(2)}$ is the Newtonian potential.]

The attitude of Refs.\ \cite{JS98,JS99,DJS1} vis \`a vis these regularization
ambiguities has been the following:  (i) one must acknowledge their existence,
because there exists, as yet, no convincingly unique extension of distribution
theory allowing one to select a preferred regularized value, and (ii) however,
one can analyze in detail the structure of these
ambiguities and show that they can be parametrized by only two (dimensionless)
parameters:  $\omega_k$ and $\omega_s$.  Indeed, after the pioneering work
\cite{JS98,JS99} which introduced these regularization ambiguity parameters, a
systematic study of the ambiguities has been conducted in the Appendix A of
\cite{DJS1} (by exploring all the possible operations by parts, as well as the
effect of having Pf$(f_1f_2) \neq$ Pf$(f_1)$ Pf$(f_2)$).  This study confirmed
the existence of only two regularization ambiguities.\footnote{Ref.\ \cite{DJS1}
made an attempt at lessening the sources of ambiguity by choosing a $D^i$ such
that the contact terms ${\cal H}^{(D)}_c$ are absent.  However, even in this
`preferred' presentation, ${\cal H}^{(D)}_f$ gave rise to the two usual ADM
ambiguities.}  As the most recent work in the ADM formalism \cite{DJS3} has
shown that the `kinetic' ambiguity $\omega_k$ was uniquely fixed by imposing
global Poincar\'e invariance, the final conclusion is, as indicated in Eq.\
(\ref{eq1}) above, that the ADM formalism introduces only {\em one}
regularization ambiguity parameter:  the `static' ambiguity $\omega_s$.

It would take us too long to explain in detail why the harmonic-coordinate
approach introduces more ambiguity parameters [four, ($s_1,s_2,r'_1,r'_2$), or
five, ($s_1,s_2,r'_1,r'_2,K$), depending on the regularization prescription,
instead of two, ($\omega_k,\omega_s$)] than the ADM one (see 
\cite{BF2,BF3,BF4}).  Let us only make a short list of the most significant
differences between the two approaches:
(i) Blanchet and Faye regularize separately many independent singular 
contributions to the spatial derivative of the gravitational field instead of 
working with the full scalar Hamiltonian as a block,
(ii) when computing their ``elementary integrals'' by analytic continuation they 
can (after contracting free indices) use the {\em ordinary} Riesz formula  
(instead of the {\em generalized} Riesz formula of \cite{JS98}, necessary to 
deal with the denominators $\propto(r_1+r_2+r_{12})^\gamma$ that appear in the 
ADM Hamiltonian),
(iii) they directly work with the full hierarchy of PN fields up to
$g_{00}=-1+\ldots+2U_8/c^8$, while the ADM approach needs to work only with the
contribution $\phi_{(6)}/c^6$ to the ``scalar'' potential,
(iv) they get two (gauge-related) ambiguities of ``logarithmic'' type 
(involving two arbitrary length scales),
and (v) they use a different coordinate system.
It would be interesting to study whether a reworking of the harmonic-coordinates
work along the more `global', and more `PN order reduced' lines of the ADM
approach would not simplify their results and get rid of several of their
ambiguities.

\section{Matching the two 3PN dynamics}

We shall now show in detail that the two 3PN dynamics are equivalent modulo a
suitable shift of particle variables.  Some time ago, Damour and Sch\"afer
\cite{DS85} studied the link, at the 2PN level, between the ADM dynamics and the
harmonic-coordinates (or DeDonder-coordinates) one.  They explicitly constructed
the map between these two descriptions of the dynamics.  Let us emphasize that
this map acts on the ``motions'', i.e.\ on the particle positions (and momenta 
or velocities) as functions of time. In other words, it gives either the
transformation (with ${\bf v}_a\equiv\dot{\bf y}_a$)
\begin{mathletters}
\label{eq6}
\begin{eqnarray}
{\bf y}_a(t) &=& {\bf Y}_a({\bf x}_b(t),{\bf p}_a(t)),
\\[2ex]
{\bf v}_a(t) &=& {\bf V}_a({\bf x}_b(t),{\bf p}_b(t)) \, ,
\end{eqnarray}
\end{mathletters}
from the ADM variables (${\bf x}_b,{\bf p}_b$) to the harmonic ones
(${\bf y}_a,{\bf v}_a$), or the inverse transformation
\begin{mathletters}
\label{eq7}
\begin{eqnarray}
{\bf x}_a(t) &=& {\bf X}_a({\bf y}_b(t), {\bf v}_b(t)),
\\[2ex]
{\bf p}_a(t) &=& {\bf P}_a({\bf y}_b(t), {\bf v}_b(t)) \, . 
\end{eqnarray}
\end{mathletters}
All variables in Eqs.\ (\ref{eq6}), (\ref{eq7}) are taken at the same value for 
their (respective) time argument. As explained in \cite{DS85} it is always 
possible to express the looked for map in this form. One has to beware that the 
transformations (\ref{eq6}) or (\ref{eq7}) are {\em not} the direct restriction 
of a coordinate transformation,
$x'^{\mu}=x^{\mu}+\xi^{\mu}(x^{\lambda},[{\bf x}_1],[{\bf x}_2])$
(where the brackets indicate {\em functional} dependence), to a field point
$x^{\mu}$ on a particle world line, but that one must take into account the
time-shift $\xi^0$ to transform the coordinate shift $\xi^i$ into the ``motion''
shift ${\bf x}'(t)-{\bf x}(t)$ [see Eqs.\ (3) of \cite{DS85}].

Among the two forms (\ref{eq6}) or (\ref{eq7}) we found that it is simplest to 
work with (\ref{eq6}). Indeed, the necessary and sufficient conditions for 
(\ref{eq6}) to map the ADM dynamics onto the harmonic one is easily seen to be 
simply
\begin{mathletters}
\begin{eqnarray}
\label{eq8}
\{{\bf Y}_a,H\} &=& {\bf V}_a,
\\[2ex]
\label{eq9}
\{{\bf V}_a,H \} &=& \left\{\{{\bf Y}_a,H\},H\right\}
= {\bf A}_a({\bf Y}_b,{\bf V}_b).
\end{eqnarray}
\end{mathletters}
All functions entering Eqs.\ (\ref{eq8}) and (\ref{eq9})
(${\bf Y}_a,H,{\bf V}_a$) are functions of the ADM phase space coordinates 
(${\bf x}_b,{\bf p}_b$). The notation $\{\cdot,\cdot\}$ denotes the usual 
Poisson bracket 
$$
\{A({\bf x}_a,{\bf p}_a),B ({\bf x}_a,{\bf p}_a)\} \equiv \sum_a \sum_i 
\left( \frac{\partial A}{\partial \, x_a^i} \, \frac{\partial B}{\partial \, 
p_{ai}} - \frac{\partial A}{\partial \, p_{ai}} \, \frac{\partial B}{\partial 
\, x_a^i} \right) \, .
$$
Finally $H$ denotes the full 3PN Hamiltonian (\ref{eq1}) while
${\bf A}_a({\bf y}_b,{\bf v}_b)$ denotes the harmonic equations of motion, Eq.\ 
(\ref{eq4}). Note that Eq. (\ref{eq8}) explicitly determines
${\bf V}_a({\bf x}_b,{\bf p}_b)$ in terms of ${\bf Y}_a({\bf x}_b,{\bf p}_b)$. 
Therefore, the problem of the mapping between $H$ and ${\bf A}_a$ is reduced to 
solving Eq.\ (\ref{eq9}) as an equation for the two unknown phase-space 
vectorial functions ${\bf Y}_1({\bf x}_b,{\bf p}_b)$ and
${\bf Y}_2({\bf x}_b,{\bf p}_b)$. We tackled this problem by the method 
of undetermined coefficients, i.e.\ by writing the most general expression for 
the PN expansion of ${\bf Y}_a({\bf x}_b,{\bf p}_b)$. We know that ${\bf Y}_a$ 
differs from ${\bf x}_a$ only at 2PN order, i.e.\
\be
\label{eq11}
{\bf Y}_a({\bf x}_b,{\bf p}_b)
= {\bf x}_a
+ \frac{1}{c^4} {\bf Y}^{\text{2PN}}_a({\bf x}_b,{\bf p}_b)
+ \frac{1}{c^6} {\bf Y}^{\text{3PN}}_a({\bf x}_b,{\bf p}_b).
\ee
The explicit expression of ${\bf Y}_a^{\text{2PN}}$ was given in Ref.\ 
\cite{DS85} (we write it here for $a=1$; the expression for $a=2$ 
being obtained by a simple relabeling $1\leftrightarrow2$):
\bea
\label{eq12}
{\bf Y}^{\text{2PN}}_1({\bf x}_a,{\bf p}_a)
&=& G m_2 \Bigg\{ \left[ \frac{5}{8} \frac{\pjpj}{m_2^2} - \frac{1}{8} 
\frac{\npj^2}{m_2^2}
+ \frac{G m_1}{r_{12}} \left(\frac{7}{4} + \frac{1}{4} \frac{m_2}{m_1}\right) 
\right]
{\bf n}_{12}
\nonumber\\[2ex]&&
\phantom{G m_2 \Bigg\{}
+ \frac{1}{2} \frac{\npj}{m_2} \frac{{\bf p}_1}{m_1}
- \frac{7}{4} \frac{\npj}{m_2} \frac{{\bf p}_2}{m_2} \Bigg\}.
\eea
Actually, as a check on the algebraic manipulation programmes (done with 
{\textsc{mathematica}}) that we wrote to solve Eq.\ (\ref{eq9}) we have 
explicitly checked that Eq.\ (\ref{eq12}) is the unique 
(translation-and-rotation-invariant) solution of the 2PN matching.

At 3PN, we write (by using translation and rotation invariance)
${\bf Y}^{\text{3PN}}_a$ in terms of some scalar functions
(here ${\bf n}_{ab}\equiv({\bf x}_a-{\bf x}_b)/r_{ab}$;
$r_{ab}\equiv\vert{\bf x}_a-{\bf x}_b\vert$)   
\bea
\label{eq13}
{\bf Y}^{\text{3PN}}_a({\bf x}_b,{\bf p}_b)
&=&  M_a\,{\bf n}_{ab} + \sum_b N_{ab}\,{\bf p}_b.
\eea
By imposing that the map reduces to the identity in the free-motion limit 
($G\to0$), it is enough to look for $M_a$ and $N_{ab}$ of the symbolic form: 
\ben
M &\propto& p^4 + \frac{p^2}{r_{12}} + \frac{1}{r_{12}^2}
+ \frac{\ln r_{12}}{r_{12}^2},
\\[2ex]
N &\propto&  p^3 + \frac{p}{r_{12}},
\een
where `$p^n$' denotes all the scalars made with 
${\bf p}_1$, ${\bf p}_2$ and ${\bf n}_{12}$ with homogeneity $p^n$, i.e.\
$$
p^n \propto \sum c_{n_1n_2n_3n_4n_5}\, \pipip^{n_1}\, \pjpjp^{n_2}\, 
\pipj^{n_3}\, \npi^{n_4}\, \npj^{n_5}
$$
with $2n_1+2n_2+2n_3+n_4+n_5=n$. We find that ${\bf Y}^{\text{3PN}}_1$ a priori 
contains 52 unknown coefficients $c_n$ (28 in $M_1$, 12 in $N_{11}$, and 12
in $N_{12}$). We did not impose any a priori constraints on the 
mass dependence of the coefficients $c_n(m_1,m_2)$ entering
${\bf Y}^{\text{3PN}}_a$.
(As a consequence we cannot make use of the $1\leftrightarrow2$ relabeling 
symmetry). Writing in full Eqs.\ (\ref{eq9}) gives a {\em linear} system of 512
equations for the $2\times52=104$ unknown coefficients $c_n$. In spite of this 
very high redundancy, we found that this system is compatible if and only if the 
arbitrary parameters $\omega_s$ and $\lambda$ are related by 
\be
\label{eq14}
\lambda = - \frac{3}{11}\omega_s - \frac{1987}{3080} \, .
\ee
Then the solution is {\em unique} and reads (for $a=1$; the solution for $a=2$ 
being obtained by relabeling $1\leftrightarrow2$)  
\bea
\label{eq15}
{\bf Y}^{\text{3PN}}_1({\bf x}_a,{\bf p}_a)
&=&  G m_2 \Bigg\{
\left[ Y^0_{1} + \frac{G m_1}{r_{12}} Y^1_1
+ \left(\frac{G m_1}{r_{12}}\right)^2 Y^2_1 \right] {\bf n}_{12}
\nonumber\\[2ex]&&
\phantom{G m_2 \Bigg\{}
+ \left( Y^0_{11}  + \frac{G m_1}{r_{12}} Y^1_{11} \right) \frac{{\bf p}_1}{m_1}
+ \left( Y^0_{12}  + \frac{G m_1}{r_{12}} Y^1_{12} \right) \frac{{\bf p}_2}{m_2} 
\Bigg\},
\eea
where
\begin{mathletters}
\label{eq16}
\bea
Y^0_{1} &=&
- \frac{1}{8} \frac{\pipj \pjpj}{m_1 m_2^3}
- \frac{1}{8} \frac{\pjpjp^2}{m_2^4}
- \frac{3}{8} \frac{\npi \npj \pjpj}{m_1 m_2^3}
+ \frac{3}{8} \frac{\npj^2 \pipj}{m_1 m_2^3}
- \frac{3}{16} \frac{\npj^2 \pjpj}{m_2^4}
\nonumber\\[2ex]&&
+ \frac{1}{8} \frac{\npi \npj^3}{m_1 m_2^3}
+ \frac{1}{16} \frac{\npj^4}{m_2^4},
\\[2ex]
Y^1_{1} &=&
  \frac{167}{48} \frac{\pipi}{m_1^2}
- \frac{105}{16} \frac{\pipj}{m_1 m_2}
+ \left( \frac{13}{6} - \frac{65}{48} \frac{m_2}{m_1} \right) 
\frac{\pjpj}{m_2^2}
- \frac{25}{48} \frac{\npi^2}{m_1^2}
+ \frac{9}{8} \frac{\npi \npj}{m_1 m_2}
\nonumber\\[2ex]&&
- \left( \frac{25}{12} - \frac{25}{48} \frac{m_2}{m_1} \right) 
\frac{\npj^2}{m_2^2},
\\[2ex]
Y^2_{1} &=&
- \frac{28387}{2520}
+ \left(\frac{49}{36} - \frac{21}{32} \pi^2\right) \frac{m_2}{m_1}
+ \frac{22}{3} \ln\frac{r_{12}}{r'_1},
\\[2ex]
Y^0_{11} &=&
- \frac{1}{4} \frac{\npj \pipi}{m_1^2 m_2}
+ \frac{\npj \pjpj}{m_2^3}
- \frac{5}{12} \frac{\npj^3}{m_2^3},
\\[2ex]
Y^1_{11} &=&
- \frac{73}{24} \frac{\npi}{m_1}
+ \left( \frac{9}{8}  - \frac{3}{2} \frac{m_2}{m_1} \right) \frac{\npj}{m_2},
\\[2ex]
Y^0_{12} &=&
- \frac{1}{8} \frac{\npi \pjpj}{m_1 m_2^2}
+ \frac{1}{4} \frac{\npj \pipj}{m_1 m_2^2}
- \frac{1}{8} \frac{\npj \pjpj}{m_2^3}
+ \frac{3}{8} \frac{\npi \npj^2}{m_1 m_2^2}
+ \frac{5}{12} \frac{\npj^3}{m_2^3},
\\[2ex]
Y^1_{12} &=&
  \frac{55}{16} \frac{\npi}{m_1}
+ \left( \frac{17}{24} + \frac{221}{48} \frac{m_2}{m_1} \right) 
\frac{\npj}{m_2}.
\eea
\end{mathletters}
The results (\ref{eq15})--(\ref{eq16}) give the explicit expression of the 
transformation
$({\bf x}_b,{\bf p}_b)\to{\bf y}_a={\bf Y}_a({\bf x}_b,{\bf p}_b)$.
To complete the knowledge of the transformation between the phase-space 
variables of the two descriptions one also needs the explicit expression of 
the transformation
$({\bf x}_b,{\bf p}_b)\to{\bf v}_a={\bf V}_a({\bf x}_b,{\bf p}_b)$.
This is straightforwardly obtained by inserting in Eq.\ (\ref{eq8}) the 
Hamiltonian of \cite{DJS3} and the results (\ref{eq11})--(\ref{eq16}) for
${\bf Y}_a({\bf x}_b,{\bf p}_b)$.
As the explicit result is very lengthy we do not display it here. [Because of 
the availability of algebraic manipulation programmes, it is safer for the 
interested reader to rederive it directly.]

Let us mention that, as a further check, we have also tried to map the 
Hamiltonian $\hiii(\omega_s,\omega_k)$ containing {\em both} $\omega_s$ and 
$\omega_k$ to ${\bf A}_a^{\text{harmonic}}$ and that we found again that the 
mapping is possible only if $\omega_k=41/24$, in complete agreement with 
\cite{DJS3}.

It should be noted that the further ambiguity parameters $r'_1$ and $r'_2$
present in the harmonic equations of motion enter our result only through some
logarithmic terms in $M_1$ [for $\ln(r_{12}/r'_1)$] and $M_2$ [for
$\ln(r_{12}/r'_2)$].  This decoupling between the two particle labels ($r'_1$
entering only ${\bf Y}^{\text{3PN}}_1$, and $r'_2$ only
${\bf Y}^{\text{3PN}}_2$) suggests (in confirmation of the discussion we gave
above) that it is not a necessity, in the harmonic approach, to introduce the
ambiguities $r'_1$ and $r'_2$.  Indeed, we see that they are locally (i.e.\
separately for each particle) introduced by the transformation of variables
between our (more ambiguity-free ADM result) and the variables defined by the
set of prescriptions of Refs.\ \cite{BF2,BF3,BF4}.  We note also that
our result confirms the finding of \cite{BF4} that $r'_1$ and $r'_2$ can be
gauged away (by a harmonicity-preserving coordinate transformation).

Anyway, the most important result is that we have shown the physical equivalence
(for invariant consequences of the dynamics) between the 3PN results of
\cite{JS98,JS99,DJS1,DJS3} of those of \cite{BF1,BF2,BF3,BF4}.  The invariants
of the 3PN dynamics depend only on {\em one} ambiguity parameter, denoted
$\omega_s$ in the ADM work, and $\lambda$ in the harmonic-coordinate one.  The
change of notation between $\omega_s$ and $\lambda$ is given in Eq.\
(\ref{eq14}) which agrees with the conclusion of \cite{BF1} which was restricted
to the circular motion case.

Could have it been different?  In view of the high redundancy of the linear
system we had to solve, it may seem quasi miraculous that the two independent
results can be made to match.  The compatibility we found is clearly a very
useful check on the algebraic computations done by both groups.  However, we
want to point out that, as both groups had already checked the global Poincar\'e
invariance of their results (see \cite{DJS3} for the ADM case, and
\cite{BF3,BF4} for the harmonic one), the possible remaining discrepancies
between the two dynamics were not very numerous.  In fact we can count precisely
the number of {\em irreducible} new coefficients entering all 3PN invariants of
a Poincar\'e-invariant dynamics.  The simplest way to do that is to use the
results of \cite{DJS2} on the 3PN ``effective one body'' dynamics \cite{BD}.
Using Poincar\'e-invariance we can reduce the dynamics to that of the {\em
relative motion}.  Following the results of \cite{DJS2} the number of
irreducible new coefficients entering the relative dynamics at the $n$PN level
is obtained by quotienting the arbitrariness in the (relative) Hamiltonian by
that in a generic (relative) canonical transformation.  This leaves only
[computing the difference between Eq.\ (3.6) and Eq.\ (3.8) of \cite{DJS2}]
\be
\label{eq19}
\left[\frac{(n+1)(n+2)}{2}+1\right] - \left[\frac{n(n+1)}{2} + 1\right] = n+1
\ee
irreducible coefficients at $n$PN. Moreover, one of the coefficients is trivial 
as it is given by the $n$PN-level expansion of the free-motion Hamiltonian 
$H_0=\sqrt{m_1^2+\pipi}+\sqrt{m_2^2+\pjpj}$. Finally, this leaves only 
$n$ non trivial irreducible coefficients at the $n$PN level, i.e., in 
particular, only three coefficients at 3PN.

For instance, in the effective one body approach,
these three coefficients are: $a_4$ [the coefficient of $(GM/R)^4$ in 
$-g_{00}^{\text{eff}}$], $d_3$ [the coefficient of $(GM/R)^3$ in 
$-g_{00}^{\text{eff}}g_{RR}^{\text{eff}}$], and $z_3$ [the coefficient of 
$P^4_R(GM/R)^2$ in the squared effective Hamiltonian]. These coefficients have 
been determined in \cite{DJS2}, from $\hiii^{\text{ADM}}$, and it was found that 
$d_3$ and $z_3$ are unambiguously determined (independently of the $\omega_s$ 
ambiguity) to be
\be
\label{eq20}
d_3 = 2(3\nu - 26)\nu, \quad \quad z_3 = 2(4-3\nu)\nu \, , 
\ee
where $\nu\equiv{m_1m_2/(m_1+m_2)^2}$ is the symmetric mass ratio, while 
$a_4$ turns out to depend on $\omega_s$:
\be
\label{eq21}
a_4 = \left( \frac{94}{3} - \frac{41}{32} \pi^2 + 2\omega_s \right)\nu \, .
\ee

In view of the sensitivity of $a_4$ to $\omega_s$, a real difference between the
ADM and the harmonic dynamics could then only have arisen as possible
discrepancies in the values of the only two unambiguous 3PN irreducible
coefficients $d_3$ and $z_3$, Eq.\ (\ref{eq20}).  As a further check, we have in
fact allowed for such differences in $d_3$ and $z_3$ by looking for the matching
of the harmonic equations of motion to a modified $\hiii^{\text{ADM}}$,
containing two extra terms corresponding to variations in both $d_3$ and $z_3$.
The result of this generalized matching was that the variations in $d_3$ and
$z_3$ had both to vanish for the matching to be possible.

\section{Conserved quantities and generalized Lagrangian in harmonic 
coordinates}

Having explicitly obtained the transformation from ADM variables to harmonic 
ones which maps the two dynamics, we can use this map to transfer all the 
useful known results of the ADM approach to the harmonic one. For instance, 
Ref.\ \cite{DJS3} has explicitly computed the ten conserved quantities of the 
binary system associated to global Poincar\'e invariance:
total energy $H({\bf x}_a,{\bf p}_a)$,
total momentum $P_i({\bf x}_a,{\bf p}_a)$,
total angular momentum $J_i({\bf x}_a,{\bf p}_a)$,
and the center-of-mass constant (boost vector)
$J^{i0}\equiv{K^i({\bf x}_a,{\bf p}_a)}
\equiv{G^i({\bf x}_a,{\bf p}_a)}-t\,P^i({\bf x}_a,{\bf p}_a)$. 
Actually, to be able to express these conserved quantities within the harmonic 
framework one needs the {\em inverse} of the transformation $(x,p)\to(y,v)$, 
i.e.\ we need to know explicitly the functions 
\begin{mathletters}
\label{eq22}
\bea
{\bf x}_a &=& {\bf X}_a({\bf y}_b,{\bf v}_b),
\\[2ex]
{\bf p}_a &=& {\bf P}_a({\bf y}_b,{\bf v}_b).
\eea
\end{mathletters}
It is just a matter of (somewhat involved) algebraic manipulations to invert 
the PN-expanded map $(x,p) \to (y,v)$ to get Eqs.\ (\ref{eq22}). By 
straightforward insertion of the formulas (\ref{eq22}), we can then (if they are 
needed) explicitly compute the following quantities in harmonic coordinates:
\begin{mathletters}
\label{hacons}
\bea
\label{eq24}
E({\bf y}_b,{\bf v}_b) &=&
H^{\text{ADM}}\biglb({\bf X}_a({\bf y}_b,{\bf v}_b),
{\bf P}_a({\bf y}_b,{\bf v}_b)\bigrb),
\\[2ex]
\label{eq25}
P_i({\bf y}_b,{\bf v}_b) &=& \sum_a P_{ai}({\bf y}_b,{\bf v}_b),
\\[2ex]
\label{eq26}
J_i({\bf y}_b,{\bf v}_b) &=& \sum_a \varepsilon_{ikl}
X_a^k({\bf y}_b,{\bf v}_b) P_{al}({\bf y}_b,{\bf v}_b),
\\[2ex]
\label{eq27}
G_i({\bf y}_b,{\bf v}_b) &=&
G_i^{\text{ADM}}\biglb({\bf X}_a({\bf y}_b,{\bf v}_b),
{\bf P}_a({\bf y}_b,{\bf v}_b)\bigrb),
\eea
\end{mathletters}
where the explicit expression of $G_i^{\text{ADM}}({\bf x}_a,{\bf p}_a)$ is 
given in Ref.\ \cite{DJS3}.  (Note that $G_i$ is not conserved but satisfies 
$dG_i/dt=P_i$.  It is a useful, and conserved, quantity in the center-of-mass
frame where $P_i=0$.)  We shall not give here the explicit expressions of
(\ref{hacons}) as modern computer means make it safer for interested readers to
do the manipulations themselves starting from the formulas we give.  We have
checked that Eq.\ (\ref{eq24}) agrees with the 3PN conserved energy obtained in
\cite{BF4}.

Our results allow us also to {\em prove} (without guess work) that the harmonic 
equations of motion derive from a {\em generalized} Lagrangian 
$L^{\text{harmonic}}({\bf y}_a,\dot{\bf y}_a,\ddot{\bf y}_a)$,
depending on positions, velocities, {\em and accelerations}. Moreover, we can 
rather simply compute the Lagrangian $L^{\text{harmonic}}$ by starting from the 
{\em phase-space} Lagrangian of our ADM Hamiltonian framework, namely
$$
L^{\text{ADM}}({\bf x}_a,\dot{\bf x}_a,{\bf p}_a)\equiv
\sum_a {\bf p}_a\cdot\dot{\bf x}_a - H^{\text{ADM}}({\bf x}_b,{\bf p}_b).
$$
Indeed, it suffices (as one easily checks) to insert the transformation
$({\bf x}_a,{\bf p}_a)\to({\bf y}_a,\dot{\bf y}_a)$ in
$L^{\text{ADM}}({\bf x}_a,\dot{\bf x}_a,{\bf p}_a)$. This gives 
\be
\label{eq28}
L^{\text{harmonic}}({\bf y}_a,\dot{\bf y}_a,\ddot{\bf y}_a)
= \sum_a {\bf P}_a({\bf y}_b,\dot{\bf y}_b) \cdot
\dot{\bf X}_a({\bf y}_b,\dot{\bf y}_b,\ddot{\bf y}_b)
- H^{\text{ADM}}\biglb({\bf X}_a({\bf y}_b,\dot{\bf y}_b),
{\bf P}_a({\bf y}_b,\dot{\bf y}_b)\bigrb).
\ee
Note that the meaning of the time derivative $\dot{X}^i_a$ in Eq.\ (\ref{eq28}) 
is 
$$
\dot{X}^i_a \equiv \frac{d X^i_a({\bf y}_b,\dot{\bf y}_b)}{dt}
\equiv \sum_b\sum_j \left( \frac{\partial{X^i_a}}{\partial{y^j_b}}\,\dot{y}^j_b
+  \frac{\partial X^i_a}{\partial{\dot{y}^j_b}}\,\ddot{y}^j_b \right).
$$
Therefore our {\em constructive} procedure for computing the harmonic Lagrangian
automatically yields a Lagrangian which is {\em linear} in the accelerations
$\ddot{\bf y}_a$.  (It was shown in \cite{DS85} that it was always possible,
for perturbatively expanded generalized Lagrangians, to reduce their
acceleration dependence to be linear.)

It should also be noted that such a linear-in-acceleration generalized
Lagrangian is not unique, but is defined only modulo the addition of
$dF({\bf y}_a,\dot{\bf y}_a)/dt$, where $F({\bf y}_a,\dot{\bf y}_a)$ is an
arbitrary\footnote{It is, however, convenient to restrict the arbitrariness in
$F$ so as to respect the symmetries of the problem:  translations, rotations,
space parity and time reversal.} scalar function of positions and velocities
(only).

When the 2PN-level generalized harmonic Lagrangian was first computed 
\cite{DD81,D83}, use was made of the addition of some
$F_{\text{2PN}}({\bf y}_a,\dot{\bf y}_a)$ to simplify (in a somewhat arbitrary 
way) the expression of $L_{\text{2PN}}({\bf y}_a,\dot{\bf y}_a,\ddot{\bf y}_a)$. 
As we are not here playing with the addition of $\dot{F}$ we should not expect
our constructive procedure (\ref{eq28}) to yield a result which coincides
with that of \cite{DD81,D83}.  We have, however, checked that our 2PN-level
result is indeed equivalent to the old one, modulo some
$dF({\bf y}_a,\dot{\bf y}_a)/dt$.  Our explicit result for the 3PN-accurate 
generalized harmonic Lagrangian reads (here ${\bf v}_a\equiv\dot{\bf y}_a$ and
${\bf a}_a\equiv\ddot{\bf y}_a$)
\be
\label{eq29}
L^{\text{harmonic}}({\bf y}_a,{\bf v}_a,{\bf a}_a)
= L_{\text{N}}({\bf y}_a,{\bf v}_a)
+ \frac{1}{c^2} L_{\text{1PN}}({\bf y}_a,{\bf v}_a)
+ \frac{1}{c^4} L_{\text{2PN}}({\bf y}_a,{\bf v}_a,{\bf a}_a)
+ \frac{1}{c^6} L_{\text{3PN}}({\bf y}_a,{\bf v}_a,{\bf a}_a).
\ee
The Newtonian and 1PN contributions to the Lagrangian (\ref{eq29}) do not depend 
on accelerations. They equal
\bea
L_{\text{N}}({\bf y}_a,{\bf v}_a)
&=& \sum_a \frac{1}{2} m_a {\bf v}_a^2
+ \frac{G m_1 m_2}{\rh},
\\[2ex]
L_{\text{1PN}}({\bf y}_a,{\bf v}_a)
&=& \frac{1}{8} m_1 \vivip^2
+ \frac{G m_1 m_2}{\rh} \left[
\frac{3}{2} \vivi - \frac{7}{4} \vivj - \frac{1}{4} \nvi \nvj
- \frac{1}{2} \frac{G m_1}{\rh} \right] + (1\leftrightarrow2).
\eea
The 2PN acceleration-dependent Lagrangian $L_{\text{2PN}}$ reads
\be
L_{\text{2PN}}({\bf y}_a,{\bf v}_a,{\bf a}_a)
= L^0_{\text{2PN}}({\bf y}_a,{\bf v}_a)
+ L^1_{\text{2PN}}({\bf y}_a,{\bf v}_a,{\bf a}_a) + (1\leftrightarrow2),
\ee
where
\begin{mathletters}
\bea
L^0_{\text{2PN}} &=& \frac{1}{16} m_1 \vivip^3
+ \frac{G m_1 m_2}{\rh} \left[
L^{01}_{\text{2PN}}
+ \frac{G m_1}{\rh} L^{02}_{\text{2PN}}
+ \bigg(\frac{G m_1}{\rh}\bigg)^2 L^{03}_{\text{2PN}} \right],
\\[2ex]
L^{01}_{\text{2PN}} &=&
  \frac{7}{8}  \vivip^2
- \frac{27}{8} \vivi \vivj
+ \frac{9}{16} \vivi \vjvj
+ \frac{15}{8} \vivj^2
+ \frac{13}{8} \nvi^2 \vivj
- \frac{5}{8}  \nvi \nvj \vivi
\nonumber\\[2ex]&&
- \frac{3}{4}  \nvi \nvj \vivj
- \frac{3}{8}  \nvj^2 \vivi
+ \frac{3}{8}  \nvi^3 \nvj
- \frac{3}{16} \nvi^2 \nvj^2,
\\[2ex]
L^{02}_{\text{2PN}} &=&
  \frac{1}{4} \vivi
- \frac{7}{4} \vivj
+ \frac{7}{4} \vjvj
+ \frac{7}{2} \nvi^2
- \frac{7}{2} \nvi \nvj
+ \frac{1}{2} \nvj^2,
\\[2ex]
L^{03}_{\text{2PN}} &=& \frac{1}{2} + \frac{19}{8} \frac{m_2}{m_1},
\\[2ex]
L^1_{\text{2PN}} &=& G m_1 m_2 \bigg[
  \frac{3}{4} \nvj \viai
- \frac{7}{4} \nvi \vjai
- \frac{7}{4} \vivj \nai
\nonumber\\[2ex]&&
\phantom{G m_1 m_2 \bigg[}
+ \frac{1}{2} \vjvj \nai
- \frac{1}{4} \nvi \nvj \nai \bigg].
\eea
\end{mathletters}
Finally, the 3PN contribution to the Lagrangian (\ref{eq29}) reads
\be
L_{\text{3PN}}({\bf y}_a,{\bf v}_a,{\bf a}_a)
= L^0_{\text{3PN}}({\bf y}_a,{\bf v}_a)
+ L^1_{\text{3PN}}({\bf y}_a,{\bf v}_a,{\bf a}_a) + (1\leftrightarrow2),
\ee
where
\begin{mathletters}
\bea
L^0_{\text{3PN}} &=& \frac{5}{128} m_1 \vivip^4
+ \frac{G m_1 m_2}{\rh} \left[
L^{01}_{\text{3PN}}
+ \frac{G m_1}{\rh} L^{02}_{\text{3PN}}
+ \bigg(\frac{G m_1}{\rh}\bigg)^2 L^{03}_{\text{3PN}}
+ \bigg(\frac{G m_1}{\rh}\bigg)^3 L^{04}_{\text{3PN}}
\right],
\\[2ex]
L^1_{\text{3PN}} &=& G m_1 m_2
\left( L^{11}_{\text{3PN}} + \frac{G m_1}{\rh} L^{12}_{\text{3PN}} \right),
\\[2ex]
L^{01}_{\text{3PN}} &=&
  \frac{11}{16} \vivip^3
- \frac{47}{16} \vivip^2 \vivj
+ \frac{25}{16} \vivip^2 \vjvj
+ \frac{23}{8}  \vivi \vivj^2
- \frac{65}{32} \vivi \vivj \vjvj
- \frac{3}{16}  \vivj^3
\nonumber\\[2ex]&&
- \frac{17}{16} \nvi \nvj \vivip^2
- \frac{3}{8}   \nvj^2      \vivip^2
+ \frac{45}{16} \nvi^2      \vivi \vivj
\nonumber\\[2ex]&&
- \frac{1}{4}   \nvi \nvj \vivi \vivj
+ \frac{11}{8}  \nvj^2      \vivi \vivj
- \frac{35}{16} \nvi^2      \vivi \vjvj
\nonumber\\[2ex]&&
+ \frac{41}{32} \nvi \nvj \vivi \vjvj
- \frac{3}{4}   \nvi^2      \vivj^2
- \frac{15}{16} \nvi \nvj \vivj^2
\nonumber\\[2ex]&&
+ \frac{19}{16} \nvi^3 \nvj   \vivi
+ \frac{9}{16}  \nvi^2 \nvj^2 \vivi
- \frac{15}{8}  \nvi \nvj^3   \vivi
+ \frac{19}{16} \nvj^4          \vivi
\nonumber\\[2ex]&&
- \frac{19}{16} \nvi^4          \vivj
-       \nvi^3 \nvj   \vivj
+ \frac{45}{32} \nvi^2 \nvj^2 \vivj
\nonumber\\[2ex]&&
- \frac{5}{16}  \nvi^5 \nvj
- \frac{5}{16}  \nvi^4 \nvj^2
+ \frac{15}{32} \nvi^3 \nvj^3,
\\[2ex]
L^{02}_{\text{3PN}} &=&
  \frac{59}{8}   \vivip^2
- \frac{199}{8}  \vivi \vivj
+ \frac{25}{3}   \vivi \vjvj
+ \frac{493}{24} \vivj^2
- \frac{113}{8}  \vivj \vjvj
+ \frac{45}{16}  \vjvjp^2
- \frac{45}{8}   \nvi^2 \vivi
\nonumber\\[2ex]&&
+ 19             \nvi \nvj \vivi
- \frac{44}{3}   \nvj^2 \vivi
- \frac{1}{4}    \nvi^2 \vivj
- \frac{61}{6}   \nvi \nvj \vivj
\nonumber\\[2ex]&&
+ \frac{67}{4}   \nvj^2 \vivj
+ \frac{275}{24} \nvi^2 \vjvj
- \frac{33}{2}   \nvi \nvj \vjvj
+ \frac{1}{4}    \nvj^2 \vjvj
- \frac{13}{3}   \nvi^4
\nonumber\\[2ex]&&
+ \frac{46}{3}   \nvi^3 \nvj
- \frac{137}{6}  \nvi^2 \nvj^2
+ \frac{34}{3}   \nvi \nvj^3,
\\[2ex]
\label{log1}
L^{03}_{\text{3PN}} &=&
  \left[ \frac{15611}{1260}
  + \left(\frac{41}{128} \pi^2 - \frac{305}{144}\right) \frac{m_2}{m_1} \right]
  \vivi
- \left[ \frac{17501}{1260}
  + \left(\frac{41}{64} \pi^2 - \frac{439}{144}\right) \frac{m_2}{m_1} \right]
  \vivj
\nonumber\\[2ex]&&
+ \left[ \frac{5}{4}
  + \left(\frac{41}{128} \pi^2 - \frac{305}{144}\right) \frac{m_2}{m_1} \right]
  \vjvj
- \left[ \frac{8243}{210}
  + \left(\frac{123}{128} \pi^2 - \frac{383}{48}\right) \frac{m_2}{m_1} \right]
  \nvi^2
\nonumber\\[2ex]&&
+ \left[ \frac{15541}{420}
  + \left(\frac{123}{64} \pi^2 - \frac{889}{48}\right) \frac{m_2}{m_1} \right]
  \nvi \nvj
+ \left[ \frac{3}{2}
  + \left(\frac{383}{48} - \frac{123}{128} \pi^2\right) \frac{m_2}{m_1} \right]  
  \nvj^2
\nonumber\\[2ex]&&
+ \left[ -\frac{22}{3} \vivi + \frac{22}{3} \vivj + 22 \nvi^2 - 22 \nvi \nvj 
\right] \ln\frac{\rh}{r'_1},
\\[2ex]
\label{log2}
L^{04}_{\text{3PN}} &=&
- \frac{3}{8}
- \left(\frac{9707}{420} + \omega_s\right) \frac{m_2}{m_1}
+ \frac{22}{3} \frac{m_2}{m_1} \ln\frac{\rh}{r'_1},
\\[2ex]
L^{11}_{\text{3PN}} &=&
- \frac{15}{4} \nvi \vivj \viai
+ \frac{5}{2}  \nvi \vjvj \viai
+ \frac{7}{4}  \nvj \vivi \viai
\nonumber\\[2ex]&&
- \frac{1}{2}  \nvj \vivj \viai
- \frac{5}{8}  \nvj \vjvj \viai
- \frac{5}{8} \nvi^2 \nvj \viai
\nonumber\\[2ex]&&
- \frac{3}{4} \nvi \nvj^2 \viai
+ \frac{5}{12} \nvj^3 \viai
- \frac{15}{8} \nvi \vivi \vjai
\nonumber\\[2ex]&&
+ \frac{1}{2}  \nvi \vivj \vjai
- \frac{1}{4}  \nvj \vivi \vjai
- \frac{1}{4}  \nvj \vivj \vjai
\nonumber\\[2ex]&&
- \frac{3}{4}  \nvi \vjvj \vjai
+ \frac{5}{12} \nvi^3       \vjai
+ \frac{3}{4} \nvi^2 \nvj \vjai
\nonumber\\[2ex]&&
- \frac{3}{8} \nvi \nvj^2 \vjai
- \frac{15}{8} \vivi \vivj \nai
+ \frac{5}{4}  \vivi \vjvj \nai
\nonumber\\[2ex]&&
+ \frac{1}{4}  \vivj^2    \nai 
- \frac{3}{4}  \vivj \vjvj \nai
+ \frac{1}{4}  \vjvjp^2    \nai
\nonumber\\[2ex]&&
- \frac{5}{8} \nvi \nvj \vivi \nai
- \frac{3}{8} \nvj^2      \vivi \nai
+ \frac{5}{4} \nvi^2      \vivj \nai
\nonumber\\[2ex]&&
+ \frac{3}{2} \nvi \nvj \vivj \nai
- \frac{3}{8} \nvj^2    \vivj \nai
- \frac{5}{4} \nvi^2    \vjvj \nai
\nonumber\\[2ex]&&
+ \frac{1}{4} \nvi \nvj \vjvj \nai
+ \frac{1}{4} \nvi^3 \nvj   \nai
+ \frac{3}{8} \nvi^2 \nvj^2 \nai
\nonumber\\[2ex]&&
- \frac{1}{8} \nvi \nvj^3   \nai,
\\[2ex]
L^{12}_{\text{3PN}} &=&
- \frac{19}{24}  \nvi \viai
- \frac{185}{24} \nvj \viai
+ \frac{205}{24} \nvi \vjai
+ \frac{67}{6}   \vivi \nai 
\nonumber\\[2ex]&&
- \frac{175}{12} \vivj     \nai
+ \frac{3}{2}    \vjvj     \nai
+ \frac{91}{24}  \nvi^2    \nai
- \frac{17}{6}   \nvi \nvj \nai
\nonumber\\[2ex]&&
- \frac{5}{4}    \nvi \viaj
+ \frac{235}{24} \nvj \viaj
- \frac{31}{3}   \nvi \vjaj
- \frac{35}{4}   \vivi \naj
\nonumber\\[2ex]&&
+ \frac{235}{24} \vivj     \naj
- \frac{21}{8}   \nvi^2    \naj
+ \frac{17}{3}   \nvi \nvj \naj.
\eea
\end{mathletters}

\section{Conclusions}

To conclude, let us emphasize that the equivalence, established in this paper,
between the existing independent approaches to 3PN dynamics is important because
it confirms the basic soundness of {\em both} approaches.  It shows that the
quite different regularization procedures devised by the two groups are
physically equivalent.  None can claim to be mathematically `better' or `more
correct' than the other one.  We have, however, pointed out that the ADM
regularization:  (i) is significantly simpler to define and apply in practice,
(ii) leads to the introduction of a minimal set of regularization ambiguities
(without extra gauge-related ambiguities).

So much for the good news.  The bad news is that having proven the physical
equivalence between the two approaches sheds no light on the problem of the
`static ambiguity' $\omega_s$.  In fact, it is sobering to note that the
enormous work which went into both 3PN investigations
\cite{JS98,JS99,DJS1,JSWeimar,DJS3,DJS2,BF1,BF2,BF3,BF4}, and which led to the
explicit evaluation of ${\cal O}(100)$ 3PN coefficients (not to mention the
${\cal O}(10^5)$ intermediate expressions which had to be computed and
manipulated) succeeded in getting only two out of the three {\em irreducible}
3PN coefficients mentioned above ($d_3$ and $z_3$; with $a_4$ staying
ambiguous).  This is all the more a pity that it was shown in \cite{DJS2} that
the 3PN-level predictions for the physically most important quantities
(dynamical behaviour \cite{DJS2}, and gravitational wave emission \cite{DIJS}
near the transition between inspiral and plunge of a binary black hole) vary
quite significantly when $\omega_s$ is allowed to vary within the plausible
range of $-10\lesssim\omega_s\lesssim10$.  This makes it urgent to further
clarify the origin of the `static ambiguity'.

Roughly speaking, it seems that this ambiguity is due to the breakdown, at
3PN, of the possibility of modelling extended (compact) objects (neutron stars
or black holes) by delta-function sources.  This is somewhat surprising because
it has been shown long ago that the extended nature of the objects (violation of
the `effacing principle') should show up only at 5PN (see Ref.\ \cite{D83},
p.\ 86).  One must probably use new techniques (or extend to 3PN-level old
techniques, such as the matching technique used in \cite{D83}) to solve this
problem.\footnote{
Let us recall that Ref.\ \cite{DJS2} found several independent arguments
suggesting that $\omega_s\simeq-9$, and maybe 
$\omega_s=-\frac{47}{3}+\frac{41}{64}\pi^2$.}
We note that it would be nice if the `effective one body' approach \cite{BD},
which is so efficient in condensing the invariant content of the dynamics to a
few coefficients, could be developed into a calculational technique, giving
explicit algorithmic recipes for directly computing the three 3PN irreducible
coefficients.

\section*{Acknowledgments}

T.D.\ thanks Laurent Schwartz for a discussion about distribution theory. 
P.J.\ and G.S.\ thank the Institut des Hautes \'Etudes Scientifiques for 
hospitality during the realization of this work.  This work was supported 
in part by the KBN Grant No.\ 2 P03B 094 17 (to P.J.).

\end{document}